\documentclass[preprint,12pt]{elsarticle} 
\usepackage{graphicx} 
\usepackage{amssymb} 
\usepackage{amsmath} 
\usepackage{amsfonts} 
\usepackage{slashed} 
\usepackage[dvipsnames]{xcolor} 

\newcommand{\Lag}{{\cal L}} 
\newcommand{\be}{\begin{equation}} 
\newcommand{\ee}{\end{equation}}

\journal{Physics Letters B} 

\begin{document}

\begin{frontmatter}

\title{$\Lambda^{\ast}(1405)$-matter: stable or unstable?}  

\author[a,b]{J.~Hrt\'{a}nkov\'{a}} 
\author[c]{N.~Barnea}
\author[c]{E.~Friedman} 
\author[c]{A.~Gal\corref{cor1}} 
\author[a]{J.~Mare\v{s}} 
\author[a,b]{M.~Sch\"{a}fer} 
\address[a]{Nuclear Physics Institute, 25068 \v{R}e\v{z}, Czech Republic} 
\address[b]{Faculty of Nuclear Sciences and Physical Engineering, \\ 
Czech Technical University in Prague, 115 19 Prague 1, Czech Republic}
\address[c]{Racah Institute of Physics, The Hebrew University, 91904 
Jerusalem, Israel} 
\cortext[cor1]{corresponding author: Avraham Gal, avragal@savion.huji.ac.il}  

\begin{abstract} 
A recent suggestion [PLB 774 (2017) 522] that 
purely-$\Lambda^{\ast}(1405)$ nuclei provide the absolute minimum energy in 
charge-neutral baryon matter for baryon-number $A\gtrsim 8$, is tested within 
RMF calculations. A broad range of $\Lambda^{\ast}$ interaction strengths, 
commensurate with $(\bar K \bar K NN)_{I=0}$ binding energy assumed to 
be of order 100~MeV, is scanned. It is found that the binding energy per 
$\Lambda^{\ast}$, $B/A$, saturates for $A\gtrsim 120$ with values of $B/A$ 
considerably below 100~MeV, implying that $\Lambda^{\ast}(1405)$ matter 
is highly unstable against strong decay to $\Lambda$ and $\Sigma$ hyperon 
aggregates. The central density of $\Lambda^{\ast}$ matter is found to 
saturate as well, at roughly twice nuclear matter density. Moreover, it is 
shown that the underlying very strong $\bar K N$ potentials, fitted for 
isospin $I=0$ to the mass and width values of $\Lambda^{\ast}(1405)$, 
fail to reproduce values of single-nucleon absorption fractions deduced across 
the periodic table from $K^-$ capture-at-rest bubble chamber experiments. 
\end{abstract} 


\end{frontmatter}

\section{Introduction} 
\label{sec:intro} 

Strangeness ($\cal S$) provides for extension of standard nuclear matter 
to strange matter in which SU(3)-octet hyperons ($\Lambda,\Sigma,\Xi$) may 
prove as abundant as nucleons~\cite{Schaffner}. Particularly interesting at 
present is the role of hyperons in the composition of the neutron star 
interior, the so called `hyperon puzzle'~\cite{CV16}. 
Little is known about the possible role of higher-mass hyperons in hadronic 
matter. However, it was recently suggested by Akaishi and Yamazaki 
(AY)~\cite{AY17} that purely-$\Lambda^{\ast}(1405)$ aggregates become 
increasingly bound with the number $A=-\cal S\,$ of $\Lambda^{\ast}$ 
constituents, reaching absolute stability for $A\gtrsim 8$. This suggestion 
for which we found no documented supporting calculations beyond $A=2$ follows 
a similar conjecture made already in 2004~\cite{YAD04}. It is worth recalling 
that solving the $A$-body Schr\"{o}dinger equation for purely attractive 
$\Lambda^{\ast}\Lambda^{\ast}$ interactions will necessarily lead to 
collapse, with the binding energy per $\Lambda^{\ast}$, $B/A$, and the 
central $\Lambda^{\ast}$ density $\rho(r\approx 0)$ diverging as $A$ increases. 
This immediately raises the question whether AY perhaps just overlooked this 
basic many-body aspect of the Schr\"{o}dinger equation in asserting that 
purely-$\Lambda^{\ast}$ matter becomes absolutely stable for some given 
value of $A$. Therefore the issue of stability has to be checked within 
calculational schemes that avoid many-body collapse. A commonly used approach 
in nuclear and hadronic physics that avoids collapse and provides sufficiently 
faithful reproduction of nuclear binding energies and densities is the 
Relativistic Mean Field (RMF) approach~\cite{SW86} which is used here. 

In this Letter, we show within RMF calculations in which strongly attractive 
$\Lambda^{\ast}\Lambda^{\ast}$ interactions are generated through scalar 
meson ($\sigma$) and vector meson ($\omega$) exchanges that both $B/A$, the 
$\Lambda^{\ast}$-matter binding energy per baryon, and the central density 
$\rho(r\approx 0)$ saturate for values of $A$ of order $A\sim 100$. For the 
case considered here, $B/A$ saturates at values between roughly 30 to 80 MeV, 
depending on details of the RMF modeling, and the associated central densities 
saturate at values about twice nuclear-matter density. This leaves $\Lambda^{
\ast}$ aggregates highly unstable against strong interaction decay governed by 
two-body conversion reactions such as $\Lambda^{\ast}\Lambda^{\ast}\to\Lambda
\Lambda,\Sigma\Sigma$. 

The plan of this note is as follows. In Sect.~\ref{sec:Kbar} we briefly 
review several few-body calculations of $\bar K$ nuclear quasibound states, 
including those based on energy independent strongly attractive $\bar K N$ 
potentials as advocated by AY, in order to introduce plausible input values 
for the $\Lambda^{\ast}\Lambda^{\ast}$ binding energy ($B_{\Lambda^{\ast}
\Lambda^{\ast}}$) used to determine the strength of the scalar and vector 
meson-exchange couplings applied in our subsequent RMF calculations. 
In Sect.~\ref{sec:atoms} we question the validity of such energy independent 
strongly attractive $\bar K N$ interactions by checking their ability 
to reproduce the single-nucleon absorption fractions deduced from $K^-$ 
capture observations in bubble chamber experiments. RMF calculations 
of purely-$\Lambda^{\ast}$ nuclei are reported in Sect.~\ref{sec:RMF}, 
showing clearly how $B/A$ and $\rho$ saturate as a function of $A$, thereby 
leaving $\Lambda^{\ast}$ matter highly unstable. A brief Conclusion section 
summarizes our results with some added discussion.

\section{$\bar K$ nuclear quasibound states} 
\label{sec:Kbar}

\begin{table}[htb]
\begin{center}
\caption{$(\bar K N)_{I=0}$, $(\bar K NN)_{I=1/2}$ and $(\bar K \bar K
NN)_{I=0}$ binding energies $B$ (in MeV) calculated using energy dependent
(E-dep.)~\cite{BGL12} and energy independent (E-indep.)~\cite{MAY13}
$\bar K N$ potentials. $(\bar K \bar K NN)_{I=0}$ binding energies are
transformed in the last row to $B_{\Lambda^{\ast}\Lambda^{\ast}}$ values.}
\vspace{5pt}
{\renewcommand{\arraystretch}{1.1}
\begin{tabular}{cccc}
\hline
$\bar K$ nuclei & (E-dep.) & (E-indep.)$_a$ & (E-indep.)$_b$  \\
\hline
$(\bar K N)_{I=0}$ & 11.4 & 26.6 & 64.2   \\
$(\bar K NN)_{I=1/2}$ & 15.7 & 51.5 & 102  \\
$(\bar K \bar K NN)_{I=0}$ & 32.1 & 93 & 190  \\
\hline
$\Lambda^{\ast}\Lambda^{\ast}$ & 9.3 & 40 & 62  \\
\hline
\end{tabular}}
\label{tab:KbarN}
\end{center}
\end{table} 

The $I=0$ antikaon-nucleon ($\bar K N$) interaction near threshold is 
attractive and sufficiently strong to form a quasibound state. Using 
a single-channel energy independent $\bar K N$ potential this quasibound 
state has been identified by AY, e.g. in Refs.~\cite{AY02,YA07}, with the 
$J^P=(1/2)^-$ $\Lambda^{\ast}(1405)$ resonance about 27~MeV below the $K^-p$ 
threshold. In contrast, in effective field theory (EFT) approaches where the 
$\bar K N$ effective single-channel potential comes out energy dependent, 
reflecting the coupling to the lower-energy $\pi\Sigma$ channel, this 
$\bar K N$ quasibound state is bound only by about 10~MeV~\cite{HJ12}. 
The difference between $\bar K N$ binding energies gets compounded in 
multi-$\bar K N$ quasibound states predicted in these two approaches, 
as demonstrated for $(\bar K \bar K NN)_{I=0}$ in Table~\ref{tab:KbarN} 
by comparing binding energies $B$ listed in the (E-dep.) column with those 
listed in the (E-indep.) columns. Regarding these two columns, we note that 
the binding energies listed in column (E-indep.)$_b$ arise by fitting the 
$(\bar K N)_{I=0}$ potential strength such that it reproduces the value 
$B(\bar K NN)_{I=1/2}=102$~MeV derived from the DISTO experiment~\cite{DISTO}. 
This derivation was challenged subsequently by the HADES 
Collaboration~\cite{HADES}. The most recent J-PARC E15~\cite{E15} dedicated 
experiment derives a value of $B(\bar K NN)_{I=1/2}$=47$\pm$3$^{+3}_{-6}$~MeV. 
Therefore, when studying energy independent $\bar K N$ potentials, we will 
keep to the (E-indep.)$_a$ scenario that also identifies the $(\bar KN)_{I=0}$ 
quasibound state with the $\Lambda^{\ast}(1405)$ resonance observed 27~MeV 
below threshold. This identification plays an essential role in the earlier 
Akaishi and Yamazaki works, Refs.~\cite{AY02,YA07}. It is worth noting that 
the more refined state-of-the-art chiral EFT approaches, with low-energy 
constants fitted to {\it all} existing $K^-p$ low-energy data, produce 
two $(\bar K N)_{I=0}$ quasibound states~\cite{PDG18}, the narrower 
and least bound of which is consistent with the (E-dep.) column of 
Table~\ref{tab:KbarN}.

\section{Kaonic atoms test} 
\label{sec:atoms} 

Here we confront the (E-indep.)$_a$ scenario of the last section with the 
broad data base of kaonic atoms which are known to provide a sensitive test 
of $\bar K N$ interaction models near threshold~\cite{FG07}. In the last 
decade several chiral EFT models of the $\bar K N$ interaction provided $K^-N$ 
scattering amplitudes based on fits to low energy $K^-p$ data, including 
kaonic hydrogen from the SIDDHARTA experiment~\cite{SIDD11,SIDD12}. Kaonic 
atom potentials based on such single-nucleon amplitudes within a sub-threshold 
kinematics approach are generally unable to fit the kaonic atom data unless 
an additional phenomenological density dependent amplitude representing 
multi-nucleon processes is introduced. In a recent work~\cite{FG17} this 
procedure was applied to several chiral EFT $\bar K N$ model amplitudes.  
Good fits to the data were reached with $\chi ^2$ values of 110 to 120 for 65 
data points. Considering that the data come from four different laboratories, 
covering the whole of the periodic table, these $\chi ^2$ values are quite 
satisfactory. This procedure was extended to include also $\bar K N$ 
amplitudes generated from the energy independent $\bar K N$ potentials used 
by Yamazaki and Akaishi (YA)~\cite{YA07} (in MeV), 
\begin{equation} 
\begin{split} 
V_{\bar K N}^{I=0}(r) =& (-595 - {\rm i} 83) \exp[-(r/0.66~{\rm fm})^2], \\ 
V_{\bar K N}^{I=1}(r) =& (-175 - {\rm i} 105) \exp[-(r/0.66~{\rm fm})^2].  
\label{eq:KN} 
\end{split} 
\end{equation} 
These potentials approximate reasonably the (E-indep.)$_a$ scenario of the 
last section. The corresponding $\bar K N$ amplitudes are shown in Fig.~15 
of Ref.~\cite{HW08}.{\footnote{We thank Tetsuo Hyodo for providing us with 
tables of these amplitudes.}} Like other models, also this model fails to 
fit kaonic atoms data on its own. Adding a phenomenological density dependent 
amplitude produces fits with $\chi ^2$ of 150 for the 65 data points, 
which is significantly inferior to fits obtained for the chiral EFT 
models considered in Ref.~\cite{FG17}. 

\begin{figure}[htb]
\begin{center}
\includegraphics[width=0.6\textwidth]{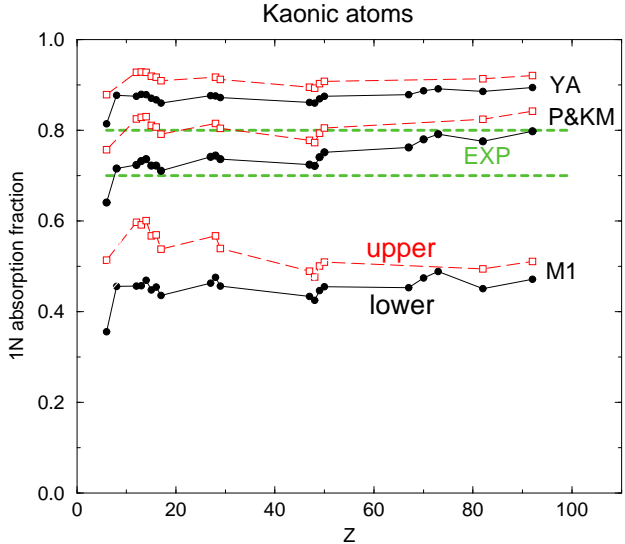}
\caption{$K^-$ single-nucleon absorption fractions calculated using $K^-N$
amplitudes from the chiral EFT models M1, P and KM, see Ref.~\cite{FG17},
and as generated from Eq.~(\ref{eq:KN}) (here marked YA). The range of
experimentally deduced fractions, 0.70--0.80, is marked by horizontal dashed
lines; see Ref.~\cite{FG17} for a comment on carbon (lowest Z points).}
\label{fig:1Nabs}
\end{center}
\end{figure}

It was shown in Ref.~\cite{FG17} that one could distinguish between different 
$\bar K N$ models by testing their ability to reproduce experimentally deduced 
values of single-nucleon absorption fractions at threshold across the periodic 
table. Fig.~\ref{fig:1Nabs} shows such fractions as calculated for four 
models of the $\bar KN$ interaction, including that of Eq.~(\ref{eq:KN}). 
Results of calculated absorptions from the so-called lower state and whenever 
provided by measured yields also from the upper state are shown for each 
kaonic atom. Experiments~\cite{DOK68,MGT71,VSW77} do not distinguish directly 
between the two types of absorption. 

As shown in the figure the $\bar K N$ interaction model of Eq.~(\ref{eq:KN}) 
(marked by YA) leads to far too large single-nucleon fractions whereas, for 
example, the Murcia (M1) model leads to too small ratios. The Kyoto-Munich 
(KM) model and the Prague (P) model, which yield predictions indistinguishable 
from each other, provide a very good agreement with experiment. The bottom 
line for the present discussion is that the $\bar K N$ interaction model of 
Eq.~(\ref{eq:KN}) does not reproduce the experimental absorption fractions.

\section{RMF calculations of purely-$\Lambda^{\ast}$ nuclei} 
\label{sec:RMF} 

Bound systems of $\Lambda^{\ast}$ hyperons are treated here in a similar 
way as applied to nuclei~\cite{SW86} and also to hypernuclei, e.g. in 
Ref.~\cite{MJ94}, within the RMF framework. In our calculations of 
$\Lambda^\ast$ nuclei, we employed the linear RMF model HS~\cite{HS81}, 
taking into account the coupling of $\Lambda^\ast$ baryons to isoscalar-scalar 
$\sigma$ and isoscalar-vector $\omega$ meson fields. Other fields considered 
in ordinary nuclei, such as isovector-vector $\vec{\rho}$ or Coulomb fields 
were disregarded since the $\Lambda^{\ast}$ is a neutral $I=0$ baryon. The 
resulting RMF model Lagrangian density for $\Lambda^{\ast}$ nuclei is of the 
form ($\hbar=c=1$ from now on): 
\begin{equation}  
\Lag = {\bar\Lambda^{\ast}}\left[\,{\rm i}\gamma^\mu D_\mu -
(M_{\Lambda^{\ast}}-g_{\sigma \Lambda^{\ast}}\sigma)\,\right]\Lambda^{\ast} + 
(\sigma,\omega_\mu\,\text{free-field terms}), 
\label{eq:Lag} 
\end{equation} 
where the covariant derivative $D_\mu=\partial_\mu+{\rm i}\,g_{\omega 
\Lambda^{\ast}}\,\omega_\mu$ couples the vector meson field $\omega$ to the 
$\Lambda^{\ast}$ baryon fields. Here we disregard the $\omega\Lambda^{\ast}$ 
tensor coupling term $f_{\omega\Lambda^{\ast}}\sigma_{\mu\nu}\omega_{\nu}$ 
which, while affecting spin-orbit splittings of single-particle levels, has 
little effect on the total binding energies of closed-shell nuclear systems 
(or $\Lambda^{\ast}$ nuclei). 

To start with, we used the HS linear model for atomic nuclei~\cite{HS81} with 
scalar and vector meson masses $m_i$ ($i=\sigma,\,\omega$) and coupling 
constants $g_{iN}$ given by 
\begin{equation} 
m_{\sigma}=520{\rm ~MeV},~~m_{\omega}=783{\rm ~MeV},~~
g_{\sigma N}=10.47,~~g_{\omega N}=13.80. 
\label{eq:LSH} 
\end{equation} 
Modifying these coupling constants in ways described below, we explored 
$\Lambda^*$ nuclei with closed shells by solving self-consistently the 
coupled system of the Klein-Gordon equations for meson fields and the 
Dirac equation for $\Lambda^{\ast}$. 

\begin{figure}[!t] 
\begin{center} 
\includegraphics[width=0.65\textwidth]{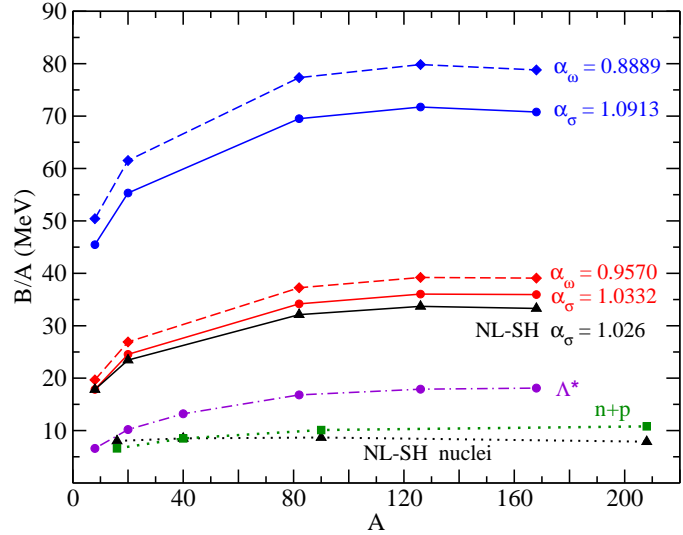}  
\caption{Binding energy of $\Lambda^{\ast}$ nuclei per $\Lambda^{\ast}$, 
$B/A$ as a function of mass number $A$, calculated within the HS and NL-SH 
RMF models for various strengths of scalar and vector fields (see text 
for details). The binding energy per nucleon in atomic nuclei is shown for 
comparison ($n+p$: HS without Coulomb and $\rho$ meson field, NL-SH nuclei: 
including these terms).} 
\label{fig:lstar} 
\end{center} 
\end{figure} 

In Fig.~\ref{fig:lstar} we show binding energy values per baryon, $B/A$, 
calculated as a function of $A$ for atomic nuclei (lowest two lines) 
and for purely $\Lambda^{\ast}$ nuclei using mostly the linear HS model. 
It is clear that $B/A$ saturates in all shown cases for $A\gtrsim 120$, 
to a value of order 10~MeV for nucleons when using parameters specified 
in Eq.~(\ref{eq:LSH}), and to a somewhat higher value in the case of 
$\Lambda^{\ast}$ nuclei (marked by $\Lambda^{\ast}$) upon using the same 
parameters. The increased $B/A$ values in this case with respect to atomic 
nuclei is due to the higher $\Lambda^{\ast}$ mass which reduces its kinetic 
energy. This is not yet the $\Lambda^{\ast}$ matter calculation we should 
pursue since when extrapolated to $A=2$ it gives a $B_{\Lambda^{\ast}
\Lambda^{\ast}}$ value of only a few MeV, whereas the calculation pursued here 
assumes a considerably stronger $\Lambda^{\ast}\Lambda^{\ast}$ binding 
corresponding to $B(\bar K \bar K NN)_{I=0}-2B(\bar K N)_{I=0}\approx 40$~MeV 
from column (E-indep.)$_a$ in Table~\ref{tab:KbarN}.{\footnote{We note for 
comparison that the scalar and vector $\Lambda^{\ast}$ couplings estimated in 
the microscopic calculations of Ref.~\cite{UHO11} within a chiral EFT model 
do not produce a bound $\Lambda^{\ast}\Lambda^{\ast}$ state.}} To renormalize 
the $\Lambda^{\ast}$ RMF calculation to such a high value of $B_{\Lambda^{
\ast}\Lambda^{\ast}}$ we need to increase $g_{\sigma N}$ or decrease 
$g_{\omega N}$ from the values listed in Eq.~(\ref{eq:LSH}). This is how the 
other $B/A$ lines marked by scaling factors $\alpha_{\sigma}$ or $\alpha_{
\omega}$ in Fig.~\ref{fig:lstar} are obtained. The appropriate values of 
$\alpha_\sigma$ and $\alpha_\omega$ are determined as follows. 

The RMF underlying attractive scalar ($\sigma$) exchange and repulsive vector 
($\omega$) exchange baryon-baryon (BB) spin-singlet $S=0$ potentials are 
given to lowest order in $(m/M)^2$ recoil corrections, disregarding tensor 
couplings, by:   
\begin{equation} 
V_{BB}(r) = g_{\omega B}^2 \,(1-\frac{3}{8}\frac{m_{\omega}^2}{M_B^2})\,
Y_{\omega}(r) - g_{\sigma B}^2 \,(1-\frac{1}{8}\frac{m_{\sigma}^2}{M_B^2})\,
Y_{\sigma}(r) 
\label{eq:DG} 
\end{equation} 
according to Dover-Gal~\cite{DG84}, or 
\begin{equation} 
V_{BB}(r) = g_{\omega B}^2 \,Y_{\omega}(r) - g_{\sigma B}^2 \, 
(1-\frac{1}{4}\frac{m_{\sigma}^2}{M_B^2})\, Y_{\sigma}(r) 
\label{eq:Mach} 
\end{equation} 
according to Machleidt~\cite{Mach88}. Here $Y_i(r)=\exp(-m_i r)/(4\pi r)$ 
is the Yukawa form for meson exchange. The difference in the $(m/M)^2$ recoil 
terms in these two forms arises from a total neglect of nonlocal contributions 
in Dover-Gal, while partially retaining them by Machleidt. Using these 
BB=$\Lambda^{\ast}\Lambda^{\ast}$ potentials, with $M_{B=\Lambda^{\ast}}=1405
$~MeV, $\Lambda^{\ast}\Lambda^{\ast}$ binding energies were calculated by 
solving a two-body Schr\"{o}dinger equation, scaling either $g_{\sigma N}$ or 
$g_{\omega M}$ according to $g_{\sigma N}\to g_{\sigma\Lambda^{\ast}}=\alpha_{
\sigma}g_{\sigma N}$ and $g_{\omega N}\to g_{\omega\Lambda^{\ast}}=\alpha_{
\omega}g_{\omega N}$ so as to get $B_{\Lambda^{\ast}\Lambda^{\ast}}=40$~MeV 
while retaining the other coupling constant fixed. The resulting scaling 
parameters are listed in Table~\ref{tab:scaling}. 

\begin{table}[htb]
\centering
\caption{Values of the scaling parameters $\alpha_{\sigma}$ and 
$\alpha_{\omega}$ for $\sigma$ and $\omega$ fields, respectively, 
each yielding $B_{\Lambda^{\ast}\Lambda^{\ast}}=40$~MeV.} 
\vspace{5pt} 
\begin{tabular}{ccc} 
\hline 
$V_{\Lambda^* \Lambda^*}$& $\alpha_{\sigma}$ & $\alpha_{\omega}$ \\ 
\hline 
Dover-Gal~(\ref{eq:DG}) & 1.0332 & 0.9750 \\
Machleidt~(\ref{eq:Mach})& 1.0913 & 0.8889 \\ 
\hline
\end{tabular}
\label{tab:scaling}
\end{table}

We then performed RMF calculations of $\Lambda^{\ast}$ nuclei using the 
renormalized coupling constants as marked to the right of each line in 
Fig.~\ref{fig:lstar}. Saturation is robust in all versions for $A\gtrsim 
120$, but the saturation value depends on which potential version is used, 
Dover-Gal (\ref{eq:DG}) or Machleidt (\ref{eq:Mach}). Scaling the $\omega$ 
meson coupling results in larger values of $\Lambda^{\ast}$ binding energies 
than by scaling the $\sigma$ meson coupling. Calculations were also 
performed using the nonlinear RMF model NL-SH~\cite{SNR93} for comparison. 
The corresponding scaling parameter $\alpha_{\sigma}=1.026$ was fitted to 
yield the binding energy of the 8$\Lambda^{\ast}$ system calculated within 
the HS model for $\alpha_{\sigma}=1.0332$. The resulting NL-SH calculation 
yields similar binding energies per $\Lambda^{\ast}$ to those produced in 
the linear HS model. Fig.~\ref{fig:lstar} clearly demonstrates that $B/A$ 
does not exceed 100 MeV in any of the versions studied here. The calculated 
values are without exception considerably lower than the $\approx 290$~MeV 
required to reduce the $\Lambda^{\ast}(1405)$ mass in the medium below that 
of the lightest hyperon $\Lambda(1116)$. This conclusion remains valid when 
$\Lambda^{\ast}$ absorption is introduced in the present RMF calculations, 
say by considering the two-body conversion processes $\Lambda^\ast\Lambda^\ast 
\rightarrow YY$ ($Y=\Lambda,~\Sigma$). Absorption normally translates into 
effective repulsion in bound state problems, thereby reducing the total 
binding energy and hence also the associated $B/A$ values in $\Lambda^\ast$ 
nuclei. 

\begin{figure}[!t]
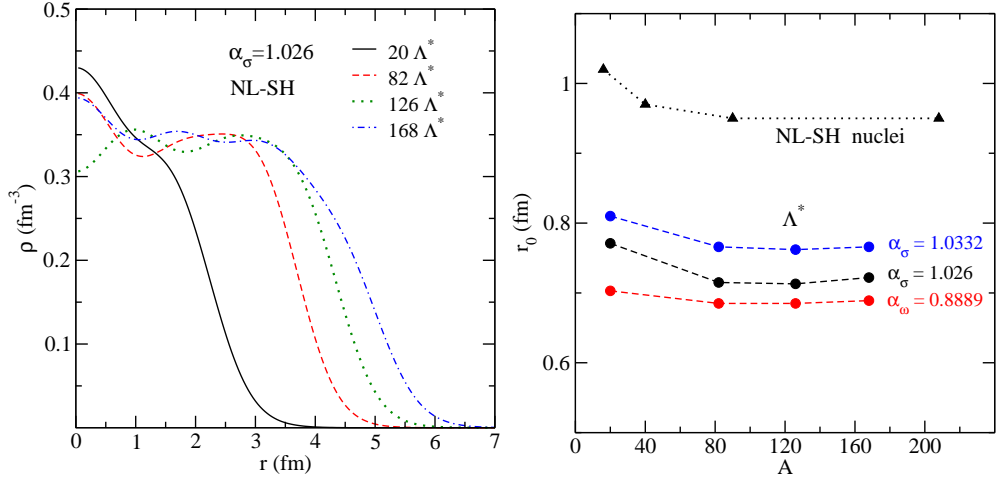

\begin{center}
\includegraphics[width=0.48\textwidth]{RMF2.eps}
\includegraphics[width=0.47\textwidth]{RMF3.eps}
\caption{Left: $\Lambda^{\ast}$ density distribution in systems composed of
20, 82, 126 and 168 $\Lambda^{\ast}$ baryons, calculated within the NL-SH RMF
model for $\alpha_\sigma=1.026$. Right: values of the r.m.s. radius parameter
$r_0$ in $\Lambda^{\ast}$ nuclei (see text) for three of the RMF models and
interaction strengths giving rise to $B/A$ lines in Fig.~\ref{fig:lstar}.
Values of $r_0$ in atomic nuclei (marked `NL-SH nuclei') calculated within 
the NL-SH model are shown for comparison.}
\label{fig:rms}
\end{center}
\end{figure}
  
Having shown that $B/A$ values saturate in $\Lambda^\ast$ nuclei to values 
less than 100~MeV, we illustrate in Fig.~\ref{fig:rms} that the central 
density $\rho(r\approx 0)$ also saturates as a function of the mass number 
$A$. This is demonstrated in the left panel for the NL-SH model and 
$\alpha_{\sigma}=1.026$. The central densities $\rho(0)$ shown in the 
figure vary in the range of 0.3--0.45~fm$^{-3}$, which is about twice nuclear 
matter density. Expressing the r.m.s. radius of the $\Lambda^\ast$ nuclear 
density distribution $\rho$ as $r_{\rm rms}=r_0\,A^{\frac{1}{3}}$, the 
variation of the radius parameter $r_0$ with $A$ is shown in the right panel 
of the figure for selected $\Lambda^\ast\Lambda^\ast$ potential versions. 
Again, the radii $r_0$ saturate with values about 0.7--0.8~fm, indicating that 
$\Lambda^\ast$ nuclei are more compressed than atomic nuclei in which $r_0$ is 
typically 0.9--1.0~fm, as shown by the upper line. The approximate constancy 
of $r_0$ with $A$ is consistent with approximately uniform $\Lambda^\ast$ 
matter density.

\section{Conclusion} 
\label{sec:concl} 

It was shown within a straightforward RMF calculation that the $\Lambda^\ast
(1405)$ stable-matter scenario promoted by AY~\cite{AY17} is unlikely to be 
substantiated in standard many-body schemes. The decisive role of Lorentz 
covariance to produce saturation in the RMF calculations of binding energies 
and sizes reported in Sect.~\ref{sec:RMF} is worth noting. Lorentz covariance 
introduces two types of baryon density, a scalar $\rho_{\rm S}={\bar B}B$ 
associated with the attractive $\sigma$ meson field and a vector $\rho_{\rm V}
={\bar B}\gamma_0 B$ associated with the repulsive $\omega$ meson field. 
Whereas $\rho_{\rm V}$ coincides with the conserved baryon density $B^\dag B$ 
(denoted simply $\rho$ on the l.h.s. of Fig.~\ref{fig:rms}), $\rho_{\rm S}$ 
shrinks with respect to $\rho_{\rm V}$ in dense matter by a multiplicative 
factor $M^\ast/E^\ast <1$, where $M^\ast=M-g_{\sigma B}\langle\sigma\rangle 
< M$ is the baryon density-dependent effective mass, thereby damping the 
attraction from the scalar $\sigma$ meson field~\cite{SW86}. Saturation in 
the RMF model is thus entirely a relativistic phenomenon. Calculations within 
the non-relativistic approach with static potentials such as (\ref{eq:DG}) 
or (\ref{eq:Mach}) would lead to collapse of systems composed of 
sufficiently large number of $\Lambda^\ast$ baryons, as it also 
holds for nucleons~\cite{Martin18}. 

Doubts were also raised in the present work on the validity of using a very 
strong and energy-independent $\bar K N$ $I=0$ dominated potential fitted 
directly to the position and width of the $\Lambda^\ast(1405)$ resonance. 
Similar potentials have been used by AY over the years to promote the case 
for strongly bound $\bar K$ nuclear clusters, see Table~\ref{tab:KbarN} here, 
and thereby also to suggest strongly attractive $\Lambda^\ast\Lambda^\ast$ 
interactions that would according to them lead to absolutely stable 
$\Lambda^\ast$ matter. It was shown in Sect.~\ref{sec:atoms} here that such 
strong and energy-independent $\bar K N$ potentials do not pass the test of 
kaonic atoms, hence casting doubts on their applicability in describing higher 
density kaonic features. Having said it, we concede that a proper description 
of high density hadronic matter, considerably beyond the $\rho\approx 2\rho_0$ 
density regime reached in our own calculations, may require the introduction 
of additional, new interaction mechanisms such as proposed recently in 
Ref.~\cite{PR15}. 

Finally, we recall related RMF calculations of multi-$\bar K$ 
nuclei~\cite{GFGM08} in which, for a given core nucleus, the resulting 
$\bar K$ separation energy $B_{\bar K}$, as well as the associated nuclear 
and $\bar K$-meson densities, were found to saturate with the number of 
$\bar K$ mesons ($\gtrsim 10$). Saturation appeared in that study robust 
against a wide range of variations, including the RMF nuclear model used 
and the type of boson fields mediating the strong interactions. In particular 
strange systems made of protons and $K^-$ mesons, as similar as possible to 
aggregates of $\Lambda^\ast(1405)$ baryons, were found in that work to be 
less bound than other strange-matter configurations. Our findings are in 
good qualitative agreement with the conclusion reached there that the SU(3)
octet hyperons ($\Lambda,~\Sigma,~\Xi$) provide, together with nucleons, for 
the lowest energy strange hadronic matter configurations~\cite{Schaffner}.

\section*{Acknowledgements} 

J.H. and M.S. acknowledge financial support from the CTU-SGS Grant No.
SGS16/243/OHK4/3T/14. The work of N.B. is supported by the Pazy Foundation 
and by the Israel Science Foundation grant No. 1308/16.

\end{document}